\documentclass[preprint1]{aastex}
\usepackage{comment}
\begin{document}

\title{Discovery of Two Small High-Velocity Compact Clouds in the Central 10 Parsecs of Our Galaxy}

\author{Shunya Takekawa$^1$, Tomoharu Oka$^{1,2}$, Yuhei Iwata$^1$, Sekito Tokuyama$^1$, and Mariko Nomura$^2$}
\affil{
$^1$School of Fundamental Science and Technology, Graduate School of Science and Technology, Keio University, 3-14-1 Hiyoshi, Yokohama, Kanagawa 223-8522, Japan\\
$^2$Department of Physics, Institute of Science and Technology, Keio University, 3-14-1 Hiyoshi, Yokohama, Kanagawa 223-8522, Japan
}
\email{shunya@z2.keio.jp}

\begin{abstract}
We discovered two small high-velocity compact clouds (HVCCs) in HCN {\it J}=4--3 and {\it J}=3--2 maps of the central 20 pc of our Galaxy.  
Both HVCCs have broad velocity widths ($\Delta V \gtrsim 40$ km s$^{-1}$) and compact sizes ($d\sim 1$ pc), and originate from the dense molecular clouds in the position--velocity space.  
One of them has a faint counterpart in a Paschen-$\alpha$ image.  
Their spatial structure, kinematics, and absence of luminous stellar object are compatible with the notion that each of the small HVCCs is driven by the plunge of an invisible compact object into a molecular cloud.  Such objects are most likely inactive, isolated black holes.  
\end{abstract}

\keywords{Galaxy: center --- galaxies: nuclei --- ISM: clouds --- ISM: molecules}

\section{Introduction}
Our Galaxy contains huge amounts of warm and dense molecular gas within 200 pc from the Galactic nucleus Sgr A$^*$.
This region is referred to as the Central Molecular Zone (CMZ; Morris \& Serabyn 1996).
Molecular gas in the CMZ show highly complex distribution and turbulent kinematics, as well as a remarkable variety of peculiar features.
In the CO {\it J}=1--0 and {\it J}=3--2 surveys of the CMZ with the Nobeyama Radio Observatory (NRO) 45 m telescope and the Atacama Submillimeter Telescope Experiment (ASTE) 10 m telescope, we discovered a mysterious population of compact ($d < 10$ pc) clouds with extremely broad velocity widths ($\Delta V \gtrsim 50$ km s$^{-1}$), called high-velocity compact clouds (HVCCs; Oka et al. 1998, 2007, 2012).
Some of the HVCCs show high CO {\it J}=3--2/{\it J}=1--0 intensity ratios ($> 1.5$) and are associated with the expanding shells and/or arcs, suggesting that they have been driven by local explosive events (e.g., Oka et al. 2001; Tanaka et al. 2007).

CO~0.02--0.02, which is one of the most energetic HVCCs, is associated with an arc-shaped emission cavity and a group of point-like infrared sources with a kinetic energy of (3--8)$\times 10^{51}$ erg (Oka et al. 2008).
This may suggest that a series of supernova explosions in an embedded massive star cluster is responsible for its formation process.
On the other hand, CO--0.40--0.22, which is another energetic HVCC, has neither expanding features nor emission cavities.
Based on the observed kinematical structure and the numerical simulations, CO--0.40--0.22 has been suggested to be a molecular cloud that has been gravitationally kicked by an inactive intermediate-mass black hole (IMBH) with $\sim10^5$ $M_\odot$ (Oka et al. 2016).

Recently, we noticed two small HVCCs in the vicinity of the circumnuclear disk (CND).
This paper reports the discovery of these small HVCCs, each of which may be driven by the high-velocity plunge of an invisible compact object.
The distance to the Galactic center is assumed to be $D = 8$  kpc.

\section{Observations}
The HCN {\it J}=4--3 (354.505 GHz) and {\it J}=3--2 (265.886 GHz) mapping observations toward the $0.15\arcdeg\times0.12\arcdeg$ area of the Galactic center were performed with the James Clerk Maxwell Telescope (JCMT) from February to July, 2016.  
The mapping area covers $-0.11\arcdeg\!\leq\!l\!\leq\!+0.04\arcdeg$ and $-0.11\arcdeg\!\leq\!b\!\leq\!+0.01\arcdeg$, including the CND, M--0.13--0.08 (+20 km s$^{-1}$ cloud), M--0.02--0.07 (+50 km s$^{-1}$ cloud), and CO 0.02--0.02 (Oka et al. 1999).  

The Heterodyne Array Receiver Program (HARP; Buckle et al. 2009) and the RxA3m receiver were used in the raster scan mode for the HCN {\it J}=4--3 and {\it J}=3--2 observations, respectively.  
As the receiver backend, the Autocorrelation Spectrometer and Imaging System (ACSIS) was operated in the 1 GHz bandwidth (488 kHz resolution) mode.
The half-power beam widths of the telescope were approximately $14''$ (350 GHz) and $20''$ (230 GHz).  
The typical system noise temperatures were $T_{\rm sys}$(SSB) $\sim 240$ K during the HARP observations and $T_{\rm sys}$(DSB) $\sim 500$ K during the RxA3m observations. 
The pointing errors were corrected every 1--1.5 hr by observing the standard sources, and the pointing accuracy was better than $2\arcsec$.  
The standard chopper-wheel method was used to calibrate the antenna temperature. 
The reference position was set to $(l, b)=(0.0\arcdeg, +0.5\arcdeg)$.  

All data were automatically reduced by the ORAC-DR pipeline software.  
The data were resampled onto $7.2\arcsec\times7.2\arcsec\times2$ km s$^{-1}$ regular grids to obtain the final maps.
The main beam efficiencies ($\eta_{\rm MB}$) were  0.64 and 0.60 at the HCN {\it J}=4--3 and {\it J}=3--2 frequencies, respectively. 
The antenna temperatures ($T_{\rm A}^*$) were converted to main beam temperatures ($T_{\rm MB}$) by multiplying with $1/\eta_{\rm MB}$.  
The rms noise levels of the resultant HCN {\it J}=4--3 and {\it J}=3--2 maps were 0.09 K and 0.15 K in $T_{\rm MB}$, respectively.

\section{Results}
The critical density of the HCN {\it J}=4--3 line is $\sim10^7$ cm$^{-3}$ and the {\it J}=4 level is 42.5 K above the rotational ground state.
The HCN {\it J}=4--3 line selectively traces the warm and dense molecular clouds in the Galactic center without absorption by colder gas in the foreground.
Figure 1(a) shows the velocity-integrated intensity map of the HCN {\it J}=4--3 line emission.
The CND, M--0.13--0.08, M--0.02--0.07, and CO 0.02--0.02 are clearly seen in the map.
At the red cross marks, ($l$, $b$)=($-0.009\arcdeg$, $-0.044\arcdeg$) and ($-0.085\arcdeg$, $-0.094\arcdeg$), we newly discovered two small HVCCs: HCN--0.009--0.044 and HCN--0.085--0.094.

\begin{figure*}[tbh]
\begin{center}
\includegraphics[width=150mm]{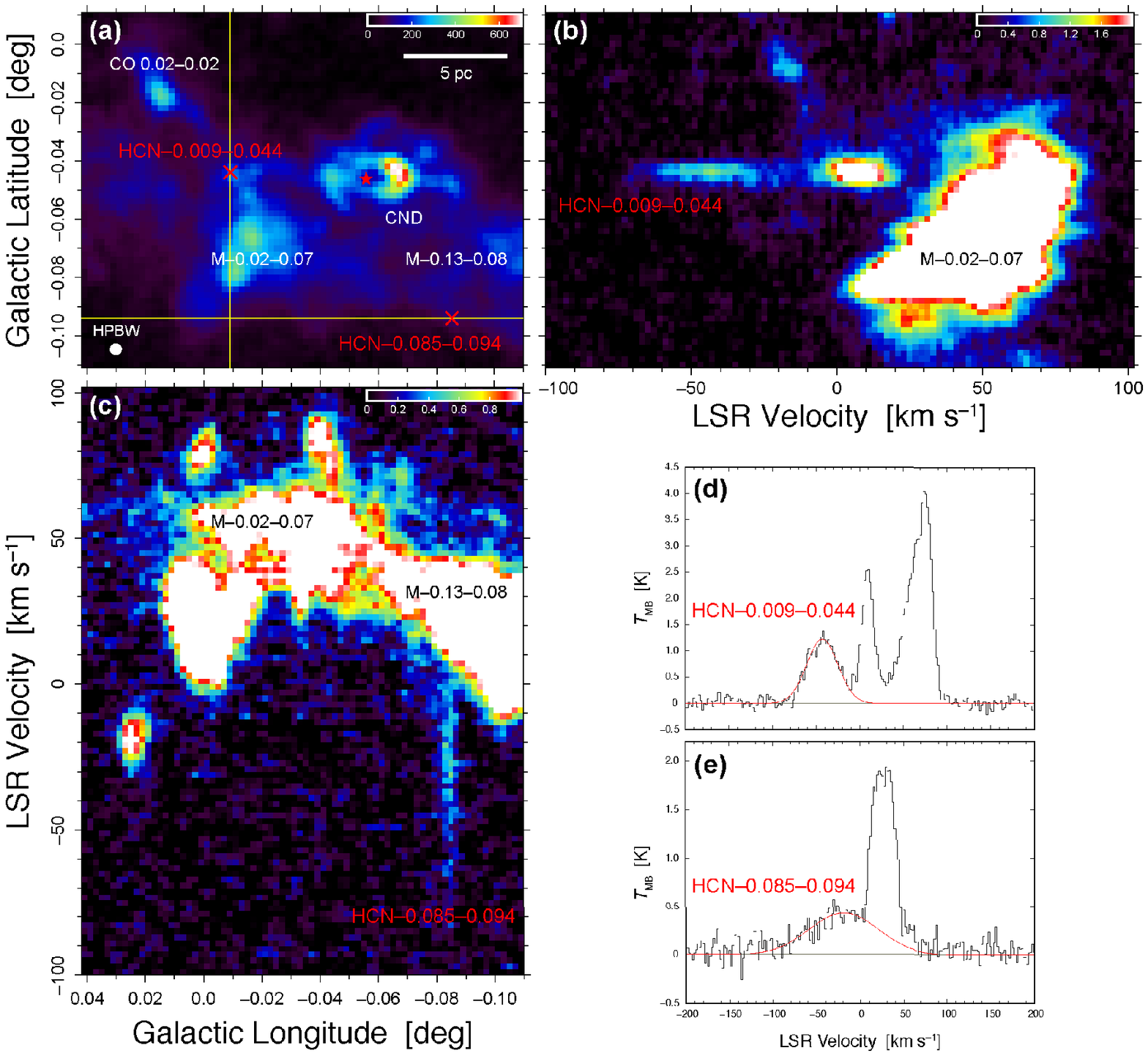}
\caption{
(a) Velocity-integrated intensity map of the HCN {\it J}=4--3 line emission.
The red star indicates the location of Sgr A$^{*}$.
The red cross marks indicate the locations of the newly found small HVCCs: HCN--0.009--0.044 and HCN--0.085--0.094.
The intensity unit is K km s$^{-1}$.
The white circle on the lower left corner indicates the HPBW ($14''$). 
(b) Latitude--velocity ($b$--$V$) map along the yellow vertical line in the panel (a). 
(c) Longitude--velocity ($l$--$V$) map along the yellow horizontal line in the panel (a).
(d) Spectrum of the HCN {\it J}=4--3 line emission at ($l$, $b$)=($-0.009\arcdeg$, $-0.044\arcdeg$).
The red curve shows the best-fitting Gaussian for the HCN--0.009--0.044 component.
(e) Spectrum of the HCN {\it J}=4--3 line emission at ($l$, $b$)=($-0.085\arcdeg$, $-0.094\arcdeg$).
The red curve shows the best-fitting Gaussian for the HCN--0.085--0.094 component.
}
\end{center}
\end{figure*}

\subsection{HCN--0.009--0.044}
Figure 1(b) shows the latitude--velocity ($b$--$V$) map at $l=-0.009\arcdeg$.
In the $b$--$V$ map, this HVCC ranges from $V_{\rm LSR}\sim -80$ to $-20$ km s $^{-1}$ and seems to stem straight from a cloud at ($b$, $V_{\rm LSR}$)$\simeq$($-0.045\arcdeg$, $+5$ km s$^{-1}$), known as the Northern Ridge, which is connected to the CND (McGary et al. 2001).
The association between HCN--0.009--0.044 and the Northern Ridge suggests that this HVCC is located within $\sim10$ pc from the Galactic nucleus.  
The HCN {\it J}=4--3 line profile peaks at $V_{\rm LSR}$ = $-40$ km s$^{-1}$ with a peak intensity of 1.3 K [Figure 1(d)].
The velocity dispersion $\sigma_{\rm V}$ was estimated to be $17.9\pm0.5$ km s$^{-1}$ from a Gaussian fitting. 
Figure 2(a) shows the velocity-integrated intensity map of HCN--0.009--0.044.
It appears as a compact clump, which is slightly elongated to the southwest, with a diameter of $\sim 2$ pc.
The size parameter given by $S=D\tan(\sqrt{\sigma_l \sigma_b})$ (Solomon et al. 1987) is 0.33 pc.

The HCN {\it J}=4--3/{\it J}=3--2 ratio suggests an excitation temperature $T_{\rm ex}$ of 22 K in the optically thin limit.
Assuming the local thermodynamic equilibrium (LTE) and a fractional abundance of [HCN]/[H$_2$] = $4.8\times10^{-8}$, which was derived for the Sgr B1 cloud (Tanaka et al. 2009; Oka et al. 2011), the total gas mass of HCN--0.009--0.044 was estimated to be $M_{\rm LTE}$ = 16 $M_{\odot}$.
The viral mass $M_{\rm VT}$ was estimated to be $2.1\times10^5$ $M_{\odot}$, which led to an extremely high viral parameter $\alpha(\equiv M_{\rm VT}/M_{\rm LTE})\sim10^4$. 
The kinetic energy $E_{\rm kin}$ was estimated to be 1.5$\times10^{47}$ erg. 
The dynamical timescale given by $\tau_{\rm d}$ = $S/\sigma_{\rm V}$ is $1.8\times10^4$ yr.
The kinetic power $L_{\rm kin}$ was estimated to be $2.6\times10^{35}$ erg s$^{-1}$ (69 $L_\odot$).
These physical parameters are summarized in Table 1.

In order to search for the counterparts in other wavelengths, we compared the HCN {\it J}=4--3 image with the 5.5 GHz radio continuum image obtained from the Jansky Very Large Array (JVLA) observations (Zhao et al. 2013, 2016), the Paschen-$\alpha$ emission image obtained from the {\it HST}/NICMOS survey (Wang et al. 2010; Dong et al. 2011), and the 0.2--8.0 keV X-ray image obtained from the {\it Chandra} observations \footnote{The {\it Chandra} images were obtained from
 \url{http//chandra.harvard.edu/photo/openFITS/xray\textunderscore data.html}} [Figure 2(b)--(d)].
Interestingly, we see a Paschen-$\alpha$ blob at the HCN emission peak position.
In the 5.5 GHz image, a filament also appears near the HCN emission peak.
Zhao et al. (2013) has noted that the 5.5 GHz filament has an X-ray counterpart CXOUGCJ174546.2--285756, which can be seen in Figure 2(d).

\subsection{HCN--0.085--0.094}
Figure 1(c) shows the longitude--velocity ($l$--$V$) map at $b=-0.094\arcdeg$.
In the $l$--$V$ map, this HVCC ranges from $V_{\rm LSR}\sim -80$ to $0$ km s$^{-1}$ and seems to stem straight from M--0.13--0.08 towards negative velocity.
Since M--0.13--0.08 is suggested to lie in the vicinity of the nucleus (Takekawa et al. 2017), this HVCC is likely to be closer to the nucleus than $\sim15$ pc.
The line profile has a peak around $V_{\rm LSR}\sim -20$ km s$^{-1}$ with a peak intensity of 0.4 K, and shows a wing-like feature [Figure 1(e)].
The best-fitting Gaussian indicates the velocity dispersion of $\sigma_{\rm V}$ = $39.5\pm4.4$ km s$^{-1}$.
Figure 3(a) shows the velocity-integrated intensity map of HCN--0.085--0.094.
It appears as a compact clump, which is slightly elongated to the south, with a diameter of $\sim 2$ pc.
The size parameter is 0.37 pc.
The physical parameters of HCN--0.085--0.094 were estimated by calculations similar to those used for HCN--0.009--0.044.
They are also listed in Table 1.

\begin{deluxetable}{ccc}
\tablewidth{0 pc}
\tablecaption{Derived Physical Parameters and Counterparts}
\tablehead{Parameters&HCN--0.009--0.044 & HCN--0.085--0.094}
\startdata
$S$  [pc] & 0.33 & 0.37 \\
$\sigma_{\rm V}$  [km s$^{-1}$] & $17.9 \pm 0.5$ & $39.5 \pm 4.4$\\
$T_{\rm ex}$  [K] &  22 & 36 \\
$M_{\rm LTE}$ [$M_{\odot}$] &16 & 13 \\
$M_{\rm VT}$  [$M_{\odot}$] & 2.1$\times10^5$ & 1.2$\times10^6$ \\
$\alpha$ & $1\times 10^4$ & $9\times10^4$\\
$E_{\rm kin}$  [erg] & $1.5\times10^{47}$& $6.0\times10^{47}$ \\
$\tau_{\rm d}$  [yr] & $1.8\times10^4$& $9.2\times10^3$\\
$L_{\rm kin}$  [erg s$^{-1}$]  & $2.6\times10^{35}$ & $2.1\times10^{36}$\\
 & (60 $L_\odot$) & (541 $L_\odot$) \\ 
\hline 
Counterparts && \\
\hline
5.5 GHz & yes & ? \\
P$\alpha$ & yes & no  \\
X-ray & yes & ? 
\enddata 
\end{deluxetable}

Figure 3(b)--(d) show the 5.5 GHz radio continuum (Zhao et al. 2013, 2016), the Paschen-$\alpha$ emission (Wang et al. 2010; Dong et al. 2011), and the 0.2--8.0 keV X-ray images.
There are diffused 5.5 GHz emissions around HCN--0.085--0.094, which are known as SE Blobs (Zhao et al. 2016).
At the south of the HVCC, there is a bright star, which emerges in the Paschen-$\alpha$ image because of the incomplete continuum subtraction.
Parts of the SE Blobs may be generated by the winds from this bright star (Zhao et al. 2016).
Except for this star, there is no Paschen-$\alpha$ source towards the HVCC.
There is no obvious X-ray counterpart, although a considerable number of sources have been detected towards the HVCC.

\begin{figure}[tbh]
\begin{center}
\includegraphics[width=85mm]{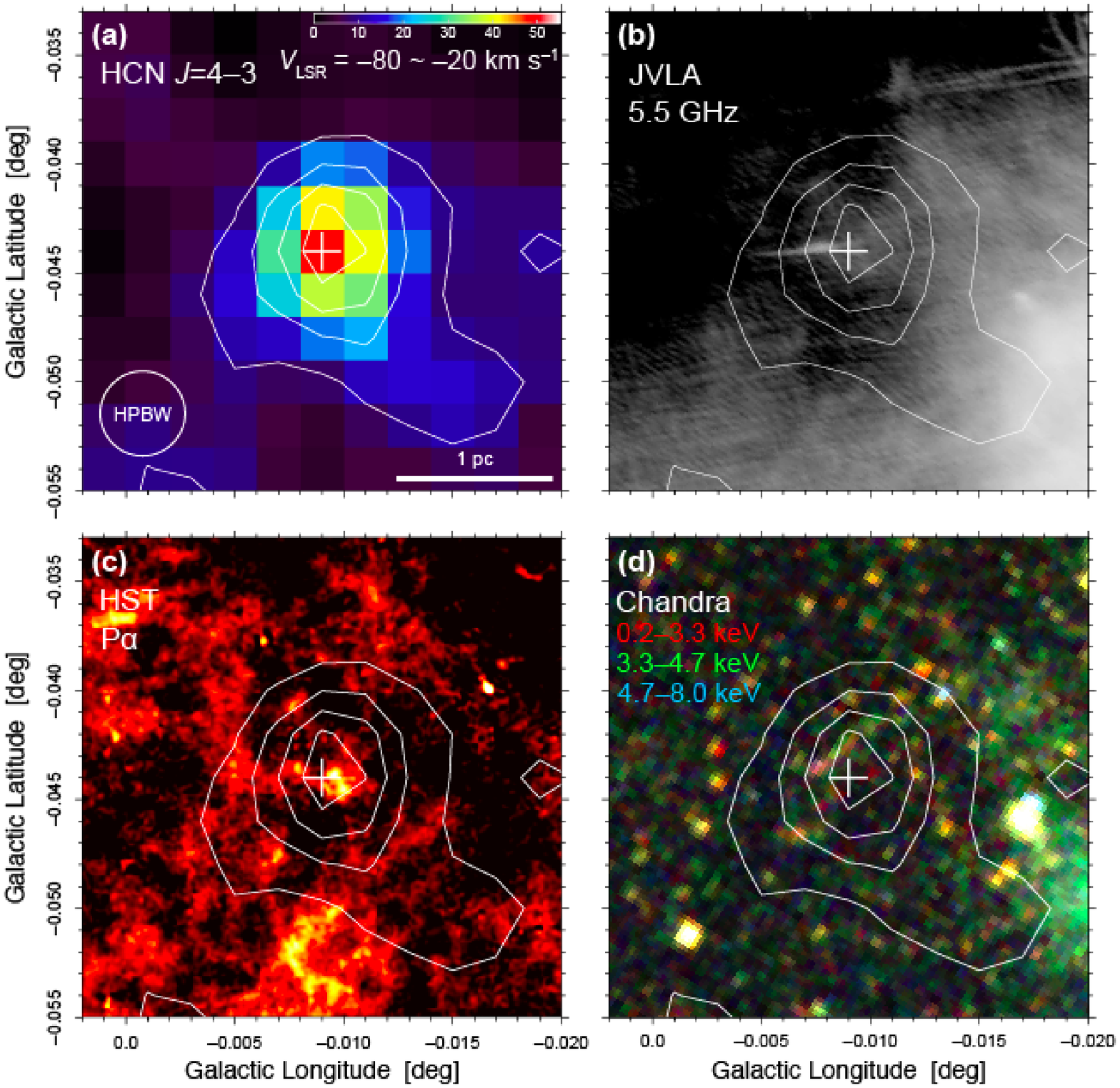}
\caption{
(a) HCN {\it J}=4--3 map of HCN--0.009--0.044 integrated over the LSR velocity from $-80$ to $-20$ km s$^{-1}$.
(b) JVLA 5.5 GHz radio continuum (Zhao et al. 2013, 2016), (c) {\it HST} Paschen-$\alpha$ emission (Wang et al. 2010; Dong et al. 2011), and (d) {\it Chandra} 3-color composite X-ray images of the same region. 
The white contours are drawn to indicate the distribution of the HVCC.
The white cross marks indicate the position of the highest velocity end.
}
\end{center}
\end{figure}

\begin{figure}[tbh]
\begin{center}
\includegraphics[width=85mm]{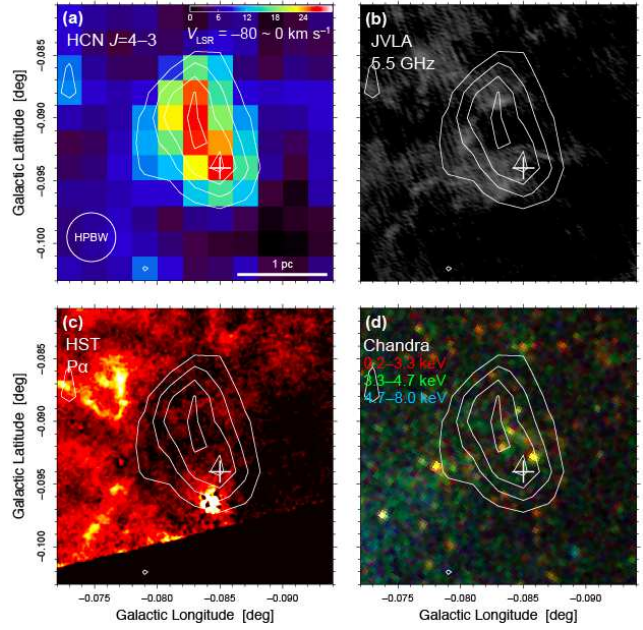}
\caption{
(a) HCN {\it J}=4--3 map of HCN--0.085--0.094 integrated over the LSR velocity from $-80$ to $0$ km s$^{-1}$.
(b)--(d) The same images as those in Figures 2 (b)--(d), but of the HCN--0.085--0.094 region.
The white contours are drawn to indicate the distribution of the HVCC.
The white cross marks indicate the position of the highest velocity end. 
}
\end{center}
\end{figure}

\begin{figure}[!tbh]
\begin{center}
\includegraphics[width=80mm]{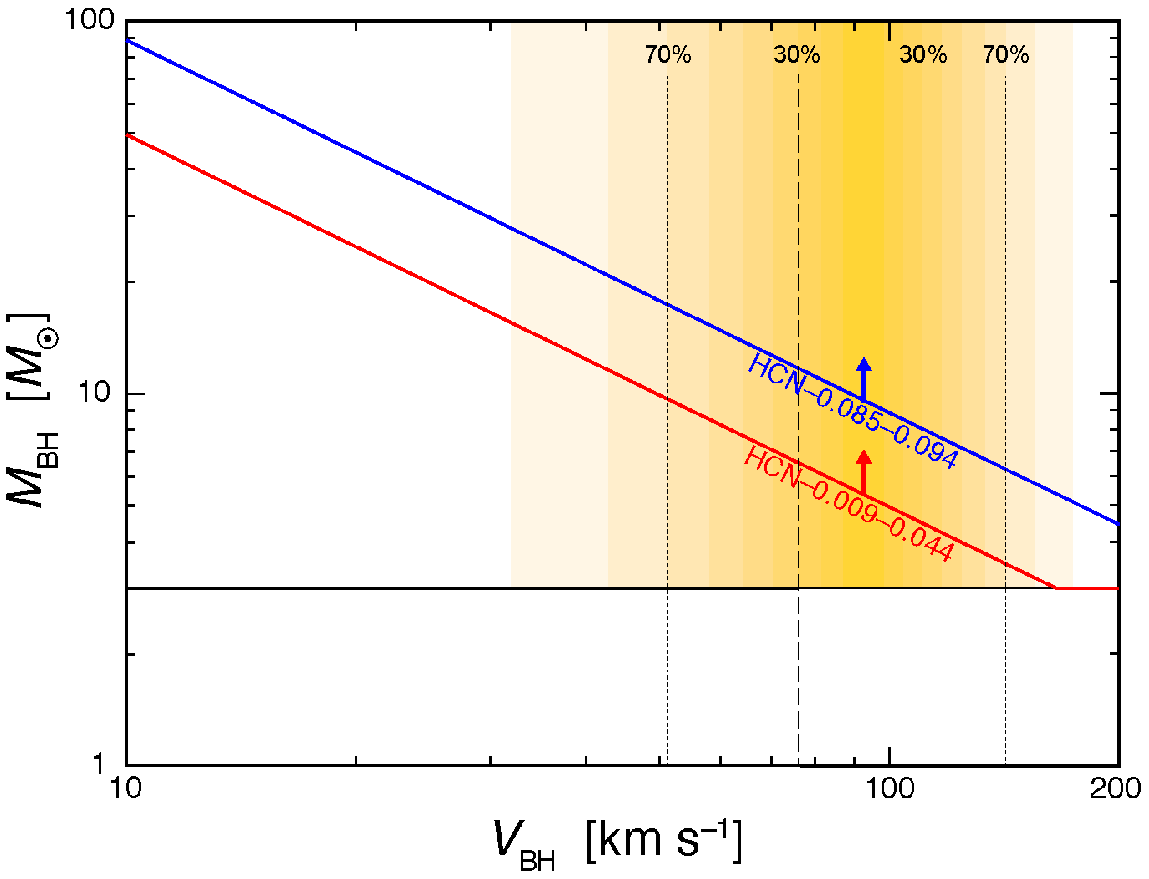}
\caption{
Lines of $P_{\rm s} = P_{\rm BH}$ for HCN--0.009--0.044 (red) and HCN--0.085--0.094 (blue).
The horizontal line indicates 3 $M_\odot$, which is approximately the Tolman--Oppenheimer--Volkoff limit.
A BH having a mass above the lines is permitted as a plunging object for the small HVCCs.
The yellow color scale and the vertical lines show the cumulative rates of the BHs from the peak velocity (92.8 km s$^{-1}$) in a gravitational potential with $r$=10 pc and $M_r$=$3\times 10^7$ $M_\odot$.
}
\end{center}
\end{figure}

\section{Discussion}
\subsection{Origin of the small HVCCs}
The sizes of the small HVCCs are smaller than those of the other previously known HVCCs, while the velocity widths are the same as that of the typical HVCCs.
Both of the small HVCCs have similar appearances and physical properties.
These facts suggest that the two HVCCs may each be formed by the same process, such as an outflow from a massive protostar, an interaction with a supernova remnant (SNR), or a cloud-cloud collision.
If the small HVCCs are driven by the outflows, there should be massive luminous ($> 10^5$ $L_\odot$) stars to supply huge kinetic energies (Maud et al. 2015).
Such lumonous stars have not been detected towards HCN--0.009--0.044, whereas there is a luminous star towards HCN--0.085--0.094 [Figure 3(c)].
This luminous star was suggested to produce stellar winds, which may be traced by the 5.5 GHz continuum (SEblb-1; Zhao et al. 2016).
However, this star is likely to lie in the foreground, because it appears in the Digitized Sky Survey (DSS) optical image without being absorbed by the foreground clouds.
The absence of luminous stellar counterparts indicate that the outflows may not be the origin of the small HVCCs.
In addition, there is no evidence of SNRs towards the small HVCCs because of the non-detection of shell-like radio continuum features.
A collision between a small cloud and a large cloud can cause a high-velocity feature like that of the small HVCCs, creating a cavity in the parent cloud (Habe \& Ohta 1992).
Each of the small HVCCs originates from a large cloud in the {\it p}--{\it V} space, and no cavity has been detected in the parent cloud.
Therefore, cloud-cloud collision may also not be the origin of the HVCCs.

Recently, we found an ultra-high-velocity feature in a giant molecular cloud (GMC) interacting with the W44 SNR (Sashida et al. 2013).
This high-velocity feature, called the Bullet (Yamada et al. 2017), has a compact appearance ($0.5 \times 0.8$ pc$^2$) and its full-width-zero-intensity velocity width exceeds 100 km s$^{-1}$.
Yamada et al. (2017) suggested that the Bullet could be driven by the high-velocity plunge of an invisible massive compact object into the dense GMC.
The kinematical structure of the Bullet can be produced by the magneto-hydrodynamical simulation of the plunging scenario (Nomura et al. in preparation).
The  {\it p}--{\it V} structures of the two small HVCCs are very similar to that of the Bullet.
Thus, each of the small HVCCs may have been driven by the high-velocity plunge of a massive compact object into a molecular cloud.
The absence of the luminous stellar counterparts for the HVCCs may suggest that the putative plunging object is an inactive BH.
The high-velocity plunge can induce dissociative shock by the sudden acceleration of gas.
The dissociative shock could have been traced by the Paschen-$\alpha$ emission at the center of HCN--0.009--0.044 [Figure 2(c)].

\subsection{BH plunging scenario}
A plunging BH gives the momentum to molecular gas, possibly driving a small HVCC.
The momentum of a small HVCC ($P_{\rm s}$) does not exceed the initial momentum of a plunging BH ($P_{\rm BH}$), i.e. $P_{\rm s} < P_{\rm BH}$.
This relation corresponds to 
\begin{eqnarray}
E_{\rm s} < E_{\rm BH}\frac{M_{\rm BH}}{M_{\rm s}}=\frac{M_{\rm BH}^2V_{\rm BH}^2}{2M_{\rm s}},
\end{eqnarray}
where $E_{\rm s}$ and $E_{\rm BH}$ are the kinetic energies,  $M_{\rm s}$ and $M_{\rm BH}$ are the masses of the small HVCC and the plunging BH, respectively, and $V_{\rm BH}$ is the relative plunging velocity of the BH.
The kinetic energy of the HVCC is estimated by $E_{\rm s} \simeq 3M_{\rm s}\sigma_{\rm V}^2/2$.
The BH mass is limited by
\begin{eqnarray}
M_{\rm BH} > \frac{\sqrt{2M_{\rm s}E_{\rm s}}}{V_{\rm BH}} = \sqrt{3}M_{\rm s}\frac{\sigma_{\rm V}}{V_{\rm BH}},
\end{eqnarray}
where $\sigma_{\rm V}$ is the velocity dispersion of the small HVCC.

Figure 4 shows the permitted ranges of  ($M_{\rm BH}$, $V_{\rm BH}$) for the two small HVCCs.
Supposing that the BHs are virialized in a gravitational potential of $\phi(r)=-GM_r/r$, the number of BHs with velocities between $v$ and $v+\Delta v$ is given by 
\begin{eqnarray}
n_{\rm BH} =  4\pi \left(\frac{3r}{2\pi G M_r}\right)^{\frac{3}{2}} v^2 \exp{\left(-\frac{3rv^2}{2GM_r}\right)}\Delta v \,N,
\end{eqnarray}
where $G$ is the gravitational constant and $N$ is the total number of BHs.
Assuming $r$=10 pc and $M_r$=$3\times 10^7$ $M_\odot$ (Sofue 2013), we infer that the peak velocity is 92.8 km s$^{-1}$.
The cumulative rates of the BHs from the peak velocity are also shown in Figure 4, which indicates that a stellar mass BH may be sufficient to produce the kinetic energy of each small HVCCs.

{We note that a hypervelocity star (Hills 1988) with a mass of $\sim1$ $M_\odot$ and a velocity of $\sim 1000$ km s$^{-1}$ also has a potential to form each of the small HVCCs.
Since such a hypervelocity star can be ejected at a rate of 10$^{-4}$ yr$^{-1}$ in the Galactic center (Yu \& Tremaine 2003), the expected number of hypervelocity stars within the central 10 pc is about unity.
On the other hand, the number of BHs within the central 10 pc was estimated to be $> 10^4$ (Antonini 2014) and the inferred number density at $r=10$ pc is $\sim 1$ $M_\odot$ pc$^{-3}$ (Merritt 2010).
Thus, the high-velocity plunge of a BH is much more likely to drive each small HVCC than that of a hypervelocity star.
}

The total number of stellar mass BHs in our Galaxy may be $\sim10^8$--$10^9$ (Agol \& Kamionkowski 2002; Caputo et al. 2017).
Most BHs would be secretly floating, being inactive and dim.
Some of them interact with molecular clouds, potentially generating high-velocity features.
Recently, we have suggested that CO--0.40--0.22, which is one of the most energetic HVCCs, may be a gravitationally kicked cloud by an inactive IMBH (Oka et al. 2016, 2017).
The HVCCs including the small HVCCs may be a key population for seeking inactive BHs hidden in molecular clouds.

\acknowledgments
The James Clerk Maxwell Telescope (JCMT) is operated by the East Asian Observatory on behalf of The National Astronomical Observatory of Japan, Academia Sinica Institute of Astronomy and Astrophysics, the Korea Astronomy and Space Science Institute, the National Astronomical Observatories of China, and the Chinese Academy of Sciences (Grant No. XDB09000000), with additional funding support from the Science and Technology Facilities Council of the United Kingdom and participating universities in the United Kingdom and Canada.
We are grateful to the JCMT staff for their excellent support during the observing run for the project M16AP058 from February to July, 2016.
This study was supported by a Grant-in-Aid for Research Fellow from the Japan Society for the Promotion of Science (15J04405).

\end{document}